\documentclass[fleqn,twoside]{article}
\usepackage{espcrc2}

\usepackage{graphicx}
\usepackage[figuresright]{rotating}

\newcommand{\AmS}{{\protect\the\textfont2
  A\kern-.1667em\lower.5ex\hbox{M}\kern-.125emS}}

\def\Journal#1#2#3#4{{#1} {\bf #2}, #3 (#4)}

\def\AP{\em Annals of Phys.}

\def\EPJC{{\em Eur. Phys. Jour.} C}
\def\JHEP{\em Jour. of High Energy Phys.}

\def\NPB{{\em Nucl. Phys.} B}
\def\PA{{\em Physica A}}
\def\PLB{{\em Phys. Lett.}  B}
\def\PRL{\em Phys. Rev. Lett.}
\def\PRD{{\em Phys. Rev.} D}
\def\ZPC{{\em Z. Phys.} C}

\newcommand{\ket}{\,\rangle}
\newcommand{\bra}{\langle \,}

\def\p{\pi}
\def\t{\tau}
\def\m{\mu}
\def\n{\nu}

\title{A Proposal for improving the Hadronization of QCD currents in TAUOLA}

\author{P. Roig\address{Instituto de F\'isica Corpuscular, IFIC, CSIC-Universitat de Val\`encia.\\ Apt. de Correus 22085, E-46071 Val\`encia, Spain}%
        \thanks{I thank the organizers of PHIPSI08 for their excellent work. I acknowledge Graziano Venanzoni and Henryk Czyz for inviting me so kindly to the Working Group on Radiative Corrections and Generators for Low Energy Hadronic Cross Section and Luminosity held after the workshop. %This work has been done in collaboration with D. G\'omez-Dumm, A. Pich and J. Portol\'es. 
I wish to thank J.~Portol\'es for a careful revision of the manuscript. P.~R. is funded by a FPU contract (MEC). This work has been supported in part by the EU MRTN-CT-2006-035482 (FLAVIAnet), by MEC (Spain) under grant FPA2007-60323, by the Spanish Consolider-Ingenio 2010 Program CPAN (CSD2007-00042) and by Generalitat Valenciana under grant GVACOMP2007-156.}}
       
\begin{document}

\begin{abstract}
\noindent After overviewing the general features of semileptonic decays of the tau lepton, I will recall the most widely used model for them, namely that of K\"uhn-Santamar\'i{}a (KS), and I will explain the subsequent works that were done along these lines and that are implemented in the TAUOLA library for analysing tau decays. After that, I will move to the description of our project that aims to achieve a theory as close as possible to QCD for the considered decays. I conclude by emphasizing the importance of the implementation of our work in TAUOLA.
\vspace{1pc}
\end{abstract}

\maketitle

\section{INTRODUCTION} \label{intro}
\noindent Hadronic decays of the $\t$ lepton are neatly splitted into the electroweak part and the strong one, responsible for the hadronization. While the first one is well under control, the second one still eludes our understanding from first principles \cite{qcd}. The division of both sectors yields a clean environment for studying the hadronization properties of QCD currents.\\
\indent The decay amplitude for the considered processes can be written as:
\begin{equation} \label{Mgraltau}
\mathcal{M}\,=\,-\frac{G_F}{\sqrt{2}}\,V_{\mathrm{CKM}}\overline{u}_{\nu_\tau}\gamma^\mu(1-\gamma_5)u_\tau \mathcal{H}_\mu\,,
\end{equation}
where the hadronic vector, $\mathcal{H}_\mu$, encodes our lack of knowledge of the precise hadronization mechanism:
\begin{equation} \label{Hmugral}
\mathcal{H}_\mu = \bra \left\lbrace  P(p_i)\right\rbrace_{i=1}^n |\left( \mathcal{V}_\mu - \mathcal{A}_\mu\right)  e^{i\mathcal{L}_{QCD}}|0\ket\,.
\end{equation}
\indent Symmetries help us to decompose $\mathcal{H}_\mu$ depending on the number of final-state mesons, $n$. In the three meson case this reads:
\begin{eqnarray} \label{Hmu3m}
\mathcal{H}_\mu = V_{1\mu} F_1^A(Q^2,s_1,s_2) + V_{2\mu} F_2^A(Q^2,s_1,s_2) +\nonumber\\
 Q_\mu F_3^A(Q^2,s_1,s_2) + i V_{3\mu} F_4^V(Q^2,s_1,s_2)\,,
\end{eqnarray}
and
\begin{eqnarray} \label{VmuQmu}
V_{1\mu} &  = & \left( g_{\mu\nu} - \frac{Q_{\mu}Q_{\nu}}{Q^2}\right) \,
(p_2 - p_1)^{\nu} ,\nonumber\\
V_{2\mu} & = & \left( g_{\mu\nu} - \frac{Q_{\mu}Q_{\nu}}{Q^2}\right) \,
(p_3 - p_1)^{\nu}\,,\nonumber\\
V_{3\mu} & = & \varepsilon_{\mu\nu\varrho\sigma}p_1^\nu\, p_2^\varrho\, p_3^{\sigma} ,\nonumber\\ Q_\mu & = & (p_1\,+\,p_2\,+\,p_3)_\mu \,,\,s_i = (Q-p_i)^2\,.
\end{eqnarray}
\indent Here $F_i$, $i=1,2,3$, correspond to the axial-vector current ($\mathcal{A}_\m$) while $F_4$ drives the vector current ($\mathcal{V}_\m$). The form factors $F_1$ and $F_2$ have a transverse structure in the total hadron momenta, $Q^\mu$, and drive a $J^P=1^+$ transition. The scalar form factor, $F_3$, vanishes with the mass of the Goldstone bosons (chiral limit) and, accordingly, gives a tiny contribution.\\
\indent It is clear that one cannot derive the $F_i$ from QCD. The aim of this communication is to convince that the most adequate way out is the use of a phenomenologically motivated theory that, in the energy region spanned by tau decays, resembles QCD as much as possible. First works on the considered modes were \cite{ks} and \cite{GGP}.\\
\indent In any case, the approximate symmetries of QCD are useful. They rule what is the theory to be used in its very low-energy domain and guide the construction of higher-energy theories, as I will sketch in section \ref{QCDMesons}. Before of that I will explain the model implemented in the MonteCarlo generators that is presently used in these decays.

\section{KS MODEL IN TAUOLA} \label{KS}
\noindent The K\"uhn-Santamar\'ia model \cite{ks} proposes phenomenologically motivated amplitudes that, by construction, fulfill the LO ($\mathcal{O}(p^2)$) $\chi$PT result in the chiral limit. However, it fails to reproduce the NLO result \cite{p03}, \cite{t3p}. It relies on the successful vector meson dominance \cite{vmd} and ensures the right behaviour at high energies, guaranteed by the Breit-Wigner propagators accounting for resonance exchange. Though widely used, its link with QCD is not proved and the proposed off-shell widths are not QFT-motivated. In any case, it provides a good description for the two and three pion decays of the tau.\\
\indent There is a lot of literature following this work studying many other hadronic decays of the $\tau$ \cite{fm}, \cite{fm2}, \cite{littau}. On the other hand, TAUOLA has grown over the years \cite{tauola} to be a complete library providing final state with full topology including neutrinos, resonances and lighter mesons and complete spin structure throughout the decay. In these works, the hadronization part of the matrix elements follows the KS model. In the remaining of the section I will focus on \cite{fm} describing the decays $\t \to KK\pi\n_\t$, $\t \to K\pi\pi\n_\t$. In it, the chiral, one- and two-resonance exchange contributions are obtained by expanding the following products of Breit-Wigner's with corresponding factors weighting the relative contribution of each resonance:
\begin{eqnarray} \label{bws}
\frac{M_{1}^2}{M_{1}^2-s-i\sqrt{s}\Gamma_{1}(s)}\,\frac{M_{2}^2}{M_{2}^2-t-i\sqrt{t}\Gamma_{2}(t)}\,.
\end{eqnarray}
\indent Throughout our work \cite{paper}, we have noticed that these studies lack from several contributions that happen to be non-negligible in our formalism, namely, they only account for $\rho^{0,-}$ and $K^{*0}$ being exchanged in one channel in the modes $K^+K^-\pi^-$, $K^-K^0\pi^0$, $K^- \pi^- \pi^+$ and $\overline{K}^0\pi^0 \pi ^-$. That is, they are only including these exchanges either in $V_1^\mu$ or $V_2^\mu$ (\ref{VmuQmu}), while one naturally obtains them in both. The authors of Refs. \cite{fm} and \cite{fm2} needed to include noticeable different masses and widths for the $\rho'$ resonance in the vector and axial-vector currents. Another point we have noticed is the inclusion of a different number of multiplets in the vector and axial-vector form factors (3 and 2, respectively). In summary, we may conclude that although the KS model describes quite well the two and three pion decay modes of the $\t$, it seems that its generalization for other three meson modes needs improvement \footnote{Current experimental situation is reviewed in section \ref{experimental situation}.}. The target of our work, whose guidelines I will be explaining in the following, is to tackle an ambitious program for providing a description as close as possible to QCD for the considered decays.
\section{QCD IN TERMS OF MESONS} \label{QCDMesons}
\noindent The key idea is to pursue a description in terms of the relevant degrees of freedom in the range of energies spanned by tau decays that incorporates the approximate symmetries of QCD therein. Well below the $\rho$(770) mass, $\chi$PT \cite{wei}, \cite{cpt} is the effective theory of QCD. Nevertheless, it only applies to a small part of the phase space in $\t$ decays \cite{col}, so one needs to include the resonances as explicit degrees of freedom. One expects those of spin one to be the most important ones, according to vector meson dominance \cite{vmd}. This emerges as a result in Resonance Chiral Theory (R$\chi$T) \cite{rcht1}, which is built upon the approximate chiral symmetry of low-energy QCD for the lightest pseudoscalar mesons and unitary symmetry for the resonances being the general procedure guaranteed by Weinberg's Theorem \cite{wei}, \cite{leut}. In this case, writing a phenomenological Lagrangian in terms of mesons fulfilling the basic symmetries of light-flavoured QCD will provide a right description. The perturbative expansion of R$\chi$T is guided by its large-$N_C$ limit \cite{lnc}. At LO in 1/$N_C$, meson dynamics is described by tree level diagrams obtained from an effective local Lagrangian including the interactions among an infinite number of stable resonances. We include finite resonance widths (a NLO effect in this expansion) within our framework \cite{width}. We are also departing from the strict $N_C\to\infty$ limit because we consider just one multiplet of resonances per set of quantum numbers (single resonance approximation \cite{sra}) and not the infinite tower predicted that cannot be included in a model independent way. For convenience, we choose to represent the spin-one mesons in the antisymmetric tensor formalism \cite{rcht2}.\\
\indent The relevant part of the R$\chi$T Lagrangian is \cite{rcht1}, \cite{vap}, \cite{vvp}, \cite{tesina}:
\begin{eqnarray} \label{Full_Lagrangian}
& & \mathcal{L}_{R\chi T}=\frac{F^2}{4}\bra u_\mu u^\mu +\chi_+ \ket+\frac{F_V}{2\sqrt{2}}\bra V_{\mu\nu} f^{\mu\nu}_+ \ket \nonumber\\
& & + \frac{i G_V}{\sqrt{2}} \bra V_{\mu\nu} u^\mu u^\nu\ket + \frac{F_A}{2\sqrt{2}}\bra A_{\mu\nu} f^{\mu\nu}_- \ket +\mathcal{L}_{\mathrm{kin}}^V\nonumber\\
& & + \mathcal{L}_{\mathrm{kin}}^A + \sum_{i=1}^{5}\lambda_i\mathcal{O}^i_{VAP} + \sum_{i=1}^7\frac{c_i}{M_V}\mathcal{O}_{VJP}^i \nonumber\\
& & + \sum_{i=1}^4d_i\mathcal{O}_{VVP}^i + \sum_{i=1}^5 \frac{g_i}{M_V} {\mathcal O}^i_{VPPP} \, ,
\end{eqnarray}
where all couplings are real, being $F$ the pion decay constant in the chiral limit. The notation is that of Ref.~\cite{rcht1}. $P$ stands for the lightest pseudoscalar mesons and $A$ and $V$ for the (axial)-vector mesons. Furthermore, all couplings in the second line are defined to be dimensionless. For the explicit form of the operators in the last line, see \cite{vap}, \cite{vvp}, \cite{tesina}.\\
\indent In order to inherit the maximum possible features of QCD, to be implemented in R$\chi$T, we still have to exploit the matching between order parameters of spontaneous chiral symmetry breaking with partonic QCD related quantities. The matching of $n$-point Green Functions in the OPE of QCD and in R$\chi$T has been shown to be a fruitful procedure \cite{rcht2}, \cite{vap}, \cite{vvp}, \cite{amo}, \cite{knecht},  \cite{consistent}. Additionally, we will demand to the vector and axial-vector form factors a Brodsky-Lepage-like behaviour \cite{brodskylepage}.\\
\indent While symmetry fully determines the structure of the operators, it is the QCD-ruled short-distance behaviour who restricts certain combinations of couplings rendering R$\chi$T predictive provided we fix these few remaining parameters restoring to phenomenology: $F_V$ could be extracted from the measured $\Gamma(\rho^0\to e^+e^-)$, $G_V$ from $\Gamma(\rho^0\to \pi^+\pi^-)$, $F_A$ from $\Gamma(a_1\to \pi\gamma)$ and the $\lambda_i$'s from $\Gamma(a_1\to \rho\pi)$ which stars on the $\t\to3\pi\n_\t$ processes themselves. $\Gamma(\omega\to \pi\gamma)$, $\Gamma(\omega\to3\pi)$ and the $\mathcal{O}(p^6)$ correction to $\Gamma(\pi \to \gamma\gamma)$ may give us information on the remaining couplings \cite{vvp}. However, we are not going to use these values but to constrain the couplings by demanding the right short-distance behaviour to the form factors appearing in $\t$ decays as we will see in section \ref{Asymptotic_behaviour}.

\section{ASYMPTOTIC BEHAVIOUR AND QCD CONSTRAINTS} \label{Asymptotic_behaviour}
\noindent There are 24 unknown couplings in $\mathcal{L}_{R\chi T}$ (\ref{Full_Lagrangian}) that may appear in the calculation of three meson decays of the $\t$.  They got reduced when computing the Feynman diagrams involved. In addition to $F_V$ and $G_V$, there only appear three combinations of the $\lbrace \lambda_i\rbrace_{i=1}^5$, four of the $\lbrace c_i\rbrace_{i=1}^7$, two of the $\lbrace d_i\rbrace_{i=1}^4$ and four of the $\lbrace g_i\rbrace_{i=1}^5$. The number of free parameters has been reduced from 24 to 15.\\
\indent We require the form factors of the $\mathcal{A}^\m$ and $\mathcal{V}^\m$ currents into $KK\p$ modes vanish at infinite transfer of momentum. As a result, we obtain constraints \cite{proc} among all axial-vector current couplings but $\lambda_0$, that are also the most general ones satisfying the demanded asymptotic behaviour in $\t\to3\p\n_\t$. Proceeding analogously with the vector current form factor results in five additional restrictions \cite{proc}. From the 24 initially free couplings in Eq. (\ref{Full_Lagrangian}), only five remain free: $c_4$, $c_1\,+\,c_2\,+\,8\,c_3\,-\,c_5$, $d_1\,+\,8\,d_2\,-\,d_3$, $g_4$ and $g_5$. After fitting $\Gamma(\omega\to3\pi)$ -using some of the relations in Refs. \cite{vvp}, \cite{Pedro}- only $c_4$ and $g_4$ remain unknown.\\
 \indent Relating the measured isovector component of $e^+e^-\to KK\pi$ to the total isovector cross-section to $KK\pi$ \cite{aleph} and employing $CVC$ to relate the latter to $\t \to KK\pi$ \cite{mal} we were able \cite{proc2} to provide a theoretical expression for $\sigma\left( e^+e^-\to KK \pi\right)$ that we fitted to BaBar data \cite{babar2} obtaining $c_4=-0.052\pm 0.003$ and $g_4=-0.20^{+0.08}_{-0.12}$ \footnote{However, we realized later that the relation used in \cite{aleph} to relate the measured isovector component in $e^+e^-\to K_S K^{\pm} \pi^{\mp}$ to $e^+e^-\to KK\pi$ is not correct. For a detailed explanation on this issue, see \cite{paper}.}. Remarkably, this procedure has allowed us to be sensitive to the sign of $c_4$ improving the work done in \cite{proc}. Our results for the decay widths of the considered channels are consistent with the PDG values \cite{PDG2006}. We plan to confront our predictions for the spectra with the forthcoming experimental data.

\section{EXPERIMENTAL SITUATION} \label{experimental situation}
\noindent CLEO \cite{cleo}, BaBar \cite{babar1} and Belle \cite{belle} have been collecting good quality data on $\t\to KK \pi\n_\t$ decays. CLEO \cite{cleo} announced that the parameterization in Ref.~\cite{fm} was unable to fit their data. They modified it by including two additional parameters in order to obtain a good fit, but at the prize of violating a property of the strong interactions, namely the Wess-Zumino \cite{WZW} normalization emanating from the chiral anomaly of QCD, as was put forward in Ref. \cite{p04}. The conclusion to be drawn from their work is clear: CLEO stated that the KS model could not fit the data and the alternative parameterization they proposed is forbidden from first principles. We would like that our expressions were used to analyse forthcoming data on the three meson decays of the $\t$. Although BaBar has not published these results yet, they have managed \cite{babar2} to split with great precission the isoscalar and isovector contributions in $e^+e^-\to KK\pi$ so that, under $CVC$ we can use it for the considered $\t$ decays, as it has been done in Ref.~\cite{mal}. We expect eagerly both BaBar and Belle new data to settle the issue of an adequate description of these decays.

\section{CONCLUSIONS}
\noindent I have reviewed the main assumptions of the KS model and its generalizations. We have found inconsistencies in the latter where the former did not fall into. Namely, the inclusion of a particle ($\rho'$) with different properties depending on where it appears, a different spectra of propagating resonances (2 or 3 multiplets) in the vector and axial-vector current form factors or the lack of several non-negligible contributions in some channels.\\
\indent I have explained the phenomenological approach that we follow that includes every relevant (and known) piece of QCD: We preserve the correct chiral limit at low energies that gives the right normalization to our form factors, we employ large-$N_C$ QCD arguments to use our theory -predicting vector meson dominance- in terms of mesons and it fulfills the high-energy conditions of the fundamental theory at the mesonic level. While we remain predictive by including only the lightest multiplet of axial-vector and vector resonances, our Lagrangian can easily be extended in a systematic way to account for the contribution of higher resonance states, whose parameters may be fitted to experiment.\\
\indent Theory and experiment are linked through event generators. The precision achieved in theoretical computations and the accuracy sought in the measurements justifies the effort that is being made in improving the MonteCarlo's. Concerning the hadronic matrix elements appearing in $\t$ decays, an estimate of the error has traditionally been obtained by comparing the results yielded by implementing the Breit-Wigner factors used in the KS model (\ref{bws}) with those incorporating resonance exchange \`a la Gounaris-Sakurai \cite{GouSak}. The inclusion of our expressions for the hadronic matrix elements in the TAUOLA \cite{tauola} library for $\t$ decays and in PHOKHARA \cite{pho} for the hadronic $e^+ e^-$ cross-section will be useful in order to describe better the hadronization of QCD currents. In the aforementioned Working Group there was unanimity in that understanding the hadronic contribution in both kinds of processes is mandatory for the MonteCarlo's employed at the $\t$-charm- and $B$-factories. This task will be of great help also in LHC and any future collider.

\end{document}